\documentstyle[psfrag,aps,preprint,epsfig,axodraw]{revtex}

\newcommand{\eqref}[1]{eq.(\ref{#1})}

\begin{document}

\tighten               

\preprint{TU-648}
\title{U(1) Flavor Symmetry and Proton Decay \\
in 
Supersymmetric Standard Model}
\author{Mitsuru Kakizaki\footnote{e-mail: kakizaki@tuhep.phys.tohoku.ac.jp}
 and Masahiro Yamaguchi\footnote{e-mail: yama@tuhep.phys.tohoku.ac.jp}}
\address{Department of Physics, Tohoku University,
Sendai 980-8578, Japan}
\date{April 2002}
\maketitle
\begin{abstract}
Issues of proton decay are examined in supersymmetric standard model with
U(1) flavor symmetry. Dimension five proton-decay operators which arise 
generically are controlled by the flavor symmetry. We show that unlike 
the minimal supersymmetric SU(5) case the proton decay modes containing 
charged lepton can have large branching ratios if the dimension five 
operators of left-handed type dominate.  Measuring the branching ratio of the 
electron mode to the muon mode may reveal the mechanism of the neutrino 
mass generation. The case with vanishing charges for Higgs doublets 
marginally survives the present experimental bound on proton life time.  
\end{abstract} 

\clearpage

\section{Introduction}

With particle contents and gauge symmetry of the standard model,
nucleon decay may take place through non-renormalizable
interaction. In fact, the gauge invariance allows dimension six
operators with baryon and lepton number violation.  They 
may be induced by exchange of X, Y gauge bosons
and colored Higgs multiplets in a grand unified theory (GUT), or may
just arise as non-renormalizable operators at a fundamental scale which we 
assume to be near the Planck scale. 
The operators of the latter origin are, however,
suppressed by the Planck mass squared, and thus the nucleon decay induced
by them is suppressed to a level where near future experiments will not
be able to reach. 

The situation is very different if one considers the
supersymmetric (SUSY) extension of the Standard Model. Existence of 
superparticles allows  gauge invariant dimension five operators which
can induce nucleon decay after superparticle dressing \cite{d5}.\footnote{
We assume $R$-parity conservation, and thus we do not consider the
most dangerous dimension four operators.}
These operators
are very dangerous because they are suppressed only by a single power of the
Planck mass. 
In fact, for the superparticle masses around 1 TeV, present proton decay
experiments constrain the mass scale of the dimension five operators much
larger than the Planck mass, or in other words, their coefficients should
be much smaller than unity when normalized by the Planck scale. 
This is indeed  embarrassing if one believes the widely accepted argument
on the generality that all operators which are
allowed by symmetry should arise with order one coefficients.  There are
many attempts to explain the smallness of these generic dimension 5
operators. They include 1) imposing some symmetry such as
the  family symmetry
\cite{Murayama:1994tc,Ben-Hamo:1994bq,Choi:1997se,Shafi:2000vm},
the discrete gauge symmetry \cite{Ibanez:1992pr},
the Peccei-Quinn symmetry\cite{Hisano:1992ne,Hisano:1995fn},
the U$(1)_A$ symmetry\cite{Hotta:1996cd}
and the  $R$ symmetry \cite{Imamura:2001es,Kurosawa:2001iq},
and 2) attributing to 
configurations of quarks and leptons in 
extra spatial dimensions\cite{Arkani-Hamed:2000dc,Kakizaki:2001ue}.

In this paper, we shall re-examine the case of U(1) flavor symmetry\cite{Froggatt:1979nt}
and closely investigate how the nucleon decays can be (or cannot be)
suppressed in the context of the minimal supersymmetric standard model
(MSSM).
Furthermore, 
we will explore the correlation between the branching fractions and 
the family structure\cite{Murayama:1994tc,Ben-Hamo:1994bq,Carone:1995xw}.

We will show that, unlike a typical SUSY SU(5) GUT, the proton decay modes
containing charged lepton can have sizable branching ratios and the ratio of 
the $\mu$ mode and the $e$ mode may give us a crucial information on the
neutrino mass generation. We will also study the proton life time. Keeping in 
mind that  
our argument based on the flavor symmetry contains uncertainties in 
coefficients of operators, we will
argue that the case with vanishing charges for the  Higgs doublets
marginally survives the present proton decay bound. 
The constraint is relaxed for negative Higgs charges. We will briefly discuss
how to generate higgsino mass $\mu$ in this case.

\section{U(1) Flavor Symmetry and Charge Assignment }
A hypothetical flavor symmetry is a symmetry to explain the Yukawa structure.
Here we consider the Froggatt-Nielsen mechanism\cite{Froggatt:1979nt} in which
we assign Froggatt-Nielsen U(1) charges $q$ to the MSSM superfields 
$Q$ and $-1$ to
the Froggatt-Nielsen field $X$. The U(1) symmetry is spontaneously broken 
when the $X$ field develops a vacuum expectation value. 
We define $\lambda$ as $\lambda \equiv \langle X \rangle/M_{\rm pl}$,
with $M_{\rm pl}$ being the Planck scale. In the  MSSM, 
we have the following superpotential
\begin{eqnarray}
  W = y_U^{ij} Q_i U^C_j H_u + y_D^{ij} Q_i D^C_j H_d + y_E^{ij} E^C_i L_j H_d,
\end{eqnarray}
where $i,j=1, 2, 3$ are generation indices. $y_U^{ij}$ {\it etc.} are Yukawa 
couplings, whose magnitudes are governed by the U(1) symmetry 
\begin{eqnarray}
  y_U^{ij} = f_U^{ij} \lambda^{q_i + u^c_j + h_u}, \quad 
  y_D^{ij} = f_D^{ij} \lambda^{q_i + d^c_j + h_d}, \quad 
  y_E^{ij} = f_E^{ij}  \lambda^{e^c_i + l_j + h_d}.
\end{eqnarray}
Here $f_U^{ij}$ {\it etc.} are somewhat arbitrary constants which are 
typically of order unity. 
In order to obtain physical masses and mixings, 
we must translate fields from a flavor basis into a mass basis 
through unitary matrices as
\begin{eqnarray}
    U_Q^{u{\rm T}} y_U U_U &=& diag(y_u, y_c, y_t), \\
    U_Q^{d{\rm T}} y_D U_D &=& diag(y_d, y_s, y_b), \\
    U_E^{\rm T} y_L U_L &=& diag(y_e, y_\mu, y_\tau),
\end{eqnarray}
from which we obtain the CKM matrix as $V_{\rm KM} = U_Q^{u \dag} U_Q^d$.

Following the conventional wisdom, 
we identify $\lambda$ with the Wolfenstein parameter $\lambda \sim 0.22$,
and determine the U($1$) charges of the MSSM fields.
The experimental values of the masses and mixings near
the Planck scale are approximated as \cite{U1}
\begin{eqnarray}
  V_{us} &\sim& \lambda, \quad V_{cb} \sim \lambda^2, 
\quad V_{ub} \sim \lambda^3,
\end{eqnarray}
and
\begin{eqnarray}
  m_u : m_c : m_t &\sim& \lambda^8 : \lambda^4 : 1, \\
  m_d : m_s : m_b &\sim& \lambda^4 : \lambda^2 : 1, \\
  m_e : m_\mu : m_\tau &\sim& \lambda^5 : \lambda^2 : 1.
\end{eqnarray}

In this paper, we will consider the following class of charge assignments:
\begin{eqnarray}
  q_1 &=& q_3 + 3, \quad q_2 = q_3 + 2, \label{eq:q}\\
  u^c_1 &=& u^c_3 + p, \quad u^c_2 = u^c_3 + 2, \label{eq:u}\\
  e^c_1 &=& e^c_3 + r - n - m, \quad e^c_2 = e^c_3 + 2 - n, \label{eq:e}\\
  d^c_1 &=& d^c_3 + 1, \quad d^c_2 = d^c_3, \label{eq:d}\\
  l_1 &=& l_3 + n + m, \quad l_2 = l_3 + n, \label{eq:l}
\end{eqnarray}
where $n,m,p$ and $r$ are integers with 
$0 \leq n \leq 2, 0 \leq m \leq 5 - n, 3 \leq p \leq 5$ and $4 \leq r \leq 5.$
Here, \eqref{eq:q} is dictated by the magnitudes of the CKM matrix elements,
\eqref{eq:u} and \eqref{eq:d} are obtained from
the up- and down-type quark mass ratios and the mixings, 
and \eqref{eq:e} and \eqref{eq:l} are determined by 
the charged lepton mass hierarchy.  Uncertainty of the coefficients $f_U^{ij}$
{\it etc.} allows some ambiguities of the charges, which are parameterized by
$p$ and $r$. 
Furthermore it follows from the fact that $m_b \sim m_\tau$ and $y_t \sim 1$
\begin{eqnarray}
  q_3 + d^c_3 = e^c_3 + l_3, \quad q_3 + u^c_3 + h_u = 0.
\end{eqnarray}

The parameters $m$ and $n$ are determined by the oscillation of
the atmospheric\cite{Fukuda:1998mi} and solar\cite{Fukuda:2001nk}
neutrinos. Here we
assume that the see-saw mechanism\cite{seesaw} 
operates to generate neutrino masses
and mixings. The overall scale of the neutrino masses is fixed by
the scale of lepton number violation, and the ratio of the masses as well
as the mixing angles are controlled by the U(1) charges. We will consider
the following two cases, both of which reproduce the bimaximal
mixing angles: \newline
i) the lopsided type\cite{lopsided}
\begin{eqnarray}
  n = 0, \quad m = 1,
\end{eqnarray}
and \newline
ii) the anarchical type\cite{Hall:2000sn,Hisano:2000wy}
\begin{eqnarray}
  n = 0, \quad m = 0,
\end{eqnarray}
with the right-handed neutrinos having appropriate U(1) charges.

To make our argument explicit,  we will  consider the following two types of 
charge assignment. 

\noindent
1) {\em best-fit} charge assignment \newline
In the first type of charge assignment, we take
\begin{eqnarray}
    p=5, \quad r=5, \quad n=0, \quad m=0~{\rm or}~1,
\end{eqnarray}
in eqs. (\ref{eq:q})--(\ref{eq:l}). We call this assignment the {\em best-fit} 
one. 
Notice that it is not consistent with 
the GUT symmetry where
the $Q$ and $U^C$, for instance, are in a single multiplet and thus 
have a common U(1)
charge. 

\noindent
2) {\em GUT-inspired} charge assignment \newline
The second type of charge assignment we will discuss is the one consistent
with the SU(5) GUT.  
Specifically we consider
\begin{eqnarray}
q_i & = & u^c_i=e^c_i, \quad q_1 = q_3+3, \quad q_2=q_3+2, \quad (p=3,~r=4), \\
d^c_i & = & l_i, \quad d^c_1 = d^c_3+1, \quad d^c_2=d^c_3 \quad (n=0,~m=1).
\end{eqnarray}
The charge assignment for $u^c$  would predict the 
up-quark Yukawa coupling of $\lambda^6$, much
larger than the actual value. 
Thus in this scheme, the small up-quark mass should be attributed to either 
accidentally small coefficients $f_U^{i1}$ of order $\lambda^2$
or the cancellation between the matrix elements.

\section{Proton Decay}
Now we  would like to discuss proton decay. 
Gauge invariance allows the dimension five operators in the
superpotential, which break
the baryon number as well as the lepton number conservation
and are divided into an LLLL part and an RRRR part:
\begin{eqnarray}
  W = \frac{1}{2M} C_L^{ijkl} Q_i Q_j Q_k L_l
  + \frac{1}{M} C_R^{ijkl} E^C_i U^C_j U^C_k D^C_l.
\end{eqnarray}
Here $C_L^{ijkl}$ and $C_R^{ijkl}$ are some coefficients, and $M$ denotes
the fundamental scale of the theory, which we identify with the Planck 
scale $M_{\rm pl}$. If these coefficient were of order unity,  they would 
cause proton decay with life time several magnitudes shorter than the present 
experimental bound. In the framework of the Froggatt-Nielsen U(1) symmetry,
we expect suppression
\begin{eqnarray}
  C_L^{ijkl} &\sim& \lambda^{q_i + q_j + q_k + l_l}
  (\sim y_U^{ij} y_D^{kl} \lambda^{-h_u - h_d} 
  \lambda^{q_j - u^c_j + l_l - d^c_l}), 
\\
  C_R^{ijkl} &\sim& \lambda^{e^c_i + u^c_j + u^c_k + d^c_l}
  (\sim y_U^{ij} y_D^{kl} \lambda^{-h_u -h_d} 
  \lambda^{e^c_i - q_i + u^c_k - q_k}).
\end{eqnarray}
Here we have omitted constants in front of the powers of $\lambda$. 
It is interesting to note here that the proton decay is suppressed if the
sum of the two Higgs charges is negative, $h_u +h_d <0$. We will come 
back to this point later on.

After wino and higgsino dressing,\footnote{
We assume that flavor mixings in the quark-squark-gluino interactions
are small enough to avoid the dangerous flavor changing processes,
and simply ignore gluino dressing diagrams in this paper.
If these generation mixings are sizable, 
the gluino dressing diagrams give considerable contributions to proton decay
\cite{Goto:1995bk}.}
we obtain proton decay amplitudes
\cite{Dimopoulos:1982dw,Ellis:1982tv,Nath:1985ub,Nath:1988tx,Hisano:1993jj,Goto:1999qg}
\footnote{
Proton decay to pion suffers from Cabbibo suppression,
  and thus we will not consider it in this paper.}
\begin{eqnarray}
  Amp(p \rightarrow K^+ \bar{\nu}_e) & \sim & \frac{f}{M} (A_1^1 + A_2^1), 
\label{eq:amp-nue}\\
  Amp(p \rightarrow K^+ \bar{\nu}_\mu) & \sim & \frac{f}{M} (A_1^2 + A_2^2), \\
  Amp(p \rightarrow K^+ \bar{\nu}_\tau) & \sim & 
  \frac{f}{M} (A_1^3 + A_2^3 + A_3 + A_4),
\label{eq:amp-nutau}
\end{eqnarray}
where
\footnote{
$C_L^{ijkl}$ in eqs. (\ref{eq:A1}) and (\ref{eq:A2}) should be replaced by
$C^{mnpq}_{L} {(U_Q^u)_m}^i {(U_Q^u)_n}^j {(U_Q^u)_p}^k {(U_L)_q}^l$.
However, since the off-diagonal elements of the mixing matrices are suppressed
by some powers of $\lambda$ (including $\lambda^0$), we deduce that
$C^{mnpq}_{L} {(U_Q^u)_m}^i {(U_Q^u)_n}^j {(U_Q^u)_p}^k {(U_L)_q}^l
\sim C_L^{ijkl}$ in this framework. The same argument is applied to $C_R$.
}
\begin{eqnarray}
  A_1^k \sim g_2^2 V_{cd}V_{cs} C_L^{221k}, \label{eq:A1}\\
  A_2^k \sim g_2^2 V_{td}V_{ts} C_L^{331k}, \label{eq:A2}\\
  A_3 \sim y_t y_\tau V_{ts} C_R^{3311}, \\
  A_4 \sim y_t y_\tau V_{td} C_R^{3312}.
\end{eqnarray}
In eqs. (\ref{eq:A1}) and (\ref{eq:A2}), we have omitted other contributions
proportional to $C_L^{112k},C_L^{113k}, \cdots$, which are of the
same order as the terms explicitly written there.
In eqs. (\ref{eq:amp-nue}) -- (\ref{eq:amp-nutau}) 
$f$ represents a loop factor, $f \sim 1/(16 \pi^2 m_{\rm SUSY})$,
where $m_{\rm SUSY}$ is a representative sparticle mass scale.
Degeneracy of the squark and slepton masses in different generations is 
implicitly assumed in the above evaluation. 
That is, we do not consider the so-called effective supersymmetry 
where the squarks and sleptons in the first two generations are very heavy.

The amplitudes of the decay $n \rightarrow K^0 \bar{\nu}$ 
are the same as those of $p \rightarrow K^+ \bar{\nu}$ up to factors of 
order  unity.

Using a numerical relation
\begin{eqnarray}
  y_t y_\tau &=& \frac{m_t m_\tau}{2 m_W^2}
  g_2^2 \left( \tan \beta + \frac{1}{\tan \beta} \right)
  \sim \frac{1}{2} \lambda^2 g_2^2 \tan \beta, \quad \tan \beta \gtrsim 3
\end{eqnarray}
at the Z-boson mass scale, we can rewrite the coefficients $A$ as 
\begin{eqnarray}
  A_1^k & \sim & A_2^k \sim g_2^2 y_t y_b 
     \lambda^{ -h_u -h_d + \delta + 8 + (l_k - l_3)}, \\
  A_3 & \sim & A_4 \sim g_2^2 y_t y_b 
     \lambda^{-h_u -h_d - \delta + 5+ p} \frac{\tan \beta}{2}, 
     \label{eq:amp}
\end{eqnarray}
where $\delta \equiv 2 q_3 - u^c_3 - e^c_3$.
Notice that the $\delta$ dependence of the LLLL operators and 
that of the RRRR operators are inverse.

As for charged lepton modes, we obtain
\begin{eqnarray}
  Amp(p \rightarrow K^0 e^+) & \sim & \frac{f}{M} A_5^1, \\
  Amp(p \rightarrow K^0 \mu^+) & \sim & \frac{f}{M} A_5^2,
\end{eqnarray}
where
\begin{eqnarray}
  A_5^k = g_2^2 \sum_i C_L^{11ik} V_{is}
  \sim g_2^2 y_t y_b \lambda^{ -h_u -h_d + \delta + 8 + (l_k - l_3)}.
\end{eqnarray}
Since the proton cannot decay to the tau lepton, 
the decay amplitudes do not contain the third generation down-type Yukawa
coupling constant.
Therefore, 
contributions from the RRRR operators to the charged lepton
modes are negligible.

\subsection{Branching Ratios}

Let us first argue the branching ratios of the proton decay.
When the LLLL operators dominate in the decay modes to the neutrinos, we find
from the above consideration
\begin{eqnarray}
   \frac{\Gamma(p \rightarrow K^0 l^+_k)}{\Gamma(p \rightarrow K^+ \bar{\nu}_k)}
  \sim O(1), \quad (k=1,2).
\end{eqnarray}
Thus, 
we predict that
\begin{eqnarray}
  \frac{\Gamma(p \rightarrow K^0 e^+)}{\Gamma(p \rightarrow K^+ \bar{\nu})}
   \sim  \lambda^{2(n+m)}, \quad
  \frac{\Gamma(p \rightarrow K^0 \mu^+)}{\Gamma(p \rightarrow K^+ \bar{\nu})} \sim \lambda^{2n},
\end{eqnarray}
where $\Gamma(p \rightarrow K^+ \bar{\nu}) = \sum_k \Gamma(p \rightarrow K^+ \bar{\nu}_k)$
is the decay width into three types of the neutrinos. In particular, for $n=0$,
which is strongly suggested by the large mixing between $\nu_{\tau}$ and
$\nu_{\mu}$, we conclude that the decay rate of the muon mode is comparable
to that of the neutrino mode. Furthermore, the decay rate to the electron
is controlled by the U(1) charge of the $l_1$, and thus we may be able to
distinguish the anarchical type assignment for the leptons from the lopsided 
one by
measuring the ratio $ \Gamma(p \rightarrow K^0 e^+) / 
\Gamma(p \rightarrow K^0 \mu^+)$.

This consequence is a striking contrast to the 
case of the minimal SU(5) SUSY GUT 
in which the proton decay operators are induced by exchanging the colored 
Higgs multiplets.
In fact, in the SU(5) case the charged lepton modes are negligibly
suppressed\cite{Ellis:1982tv,Nath:1985ub}. 
This is because the decay amplitudes into the charged leptons
are proportional to the very small up-quark mass, 
while those into the neutrinos can be proportional to the 
charm- or top-quark mass.

On the other hand, when the RRRR operators are dominant, the proton mainly
decays through
$p \to K^+ \bar \nu_{\tau}$. Thus 
we expect
\begin{eqnarray}
   \frac{\Gamma(p \rightarrow K^0 e^+, K^0 \mu^+)}
  {\Gamma(p \rightarrow K^+ \bar{\nu})} \ll O(1).
\end{eqnarray}

Which of the operators, LLLL or RRRR, dominates depends on the charges one
assumes.
When we take the {\em best-fit} charge assignment, $l_2-l_3=0$ and 
$p=5$, whereas $\delta$ is a parameter which is not fixed.
Then, for
\begin{equation}
  \delta \lesssim 0.9 - 0.3 \ln(\tan \beta/3),
\end{equation}
the LLLL contributions dominate over the RRRR ones.

In the GUT-inspired case, $l_2 - l_3 = 0, p = 3$ and $\delta = 0$.\footnote{
If the theory is really embedded into a GUT group, 
one has to take into account 
the contribution from the colored Higgs exchange.
In the GUT framework, our argument given here is valid when 
some mechanism operates to  suppress the proton 
decays mediated by the colored Higgses and thus the genuine dimension
five operators controlled by the flavor symmetry are the dominant sources of 
the proton decay. The suppression mechanisms of the proton decay operators 
coming from the exchange of the colored Higgses have been proposed, for
instance, in Refs.
\cite{Hisano:1995fn,Hotta:1996cd,Maekawa:2001uk,Altarelli:2000fu}. See also 
Ref. \cite{Babu:2002fs} 
and references therein.} 
It follows  from \eqref{eq:amp} that
for small $\tan \beta$, the LLLL amplitudes are of the same order as
the RRRR amplitudes.
Thus, we expect that the muon mode is comparable to the
neutrino mode and the electron mode is suppressed by $\lambda^2$. Namely, 
\begin{eqnarray}
  \frac{\Gamma(p \rightarrow K^0 \mu^+)}{\Gamma(p \rightarrow K^+ \bar{\nu})}
  \sim 1, \quad 
  \frac{\Gamma(p \rightarrow K^0 e^+)}{\Gamma(p \rightarrow K^+ \bar{\nu})} 
 \sim \lambda^2 \quad (\tan \beta \sim 3).
\end{eqnarray}
On the contrary, for large $\tan \beta$, 
the RRRR contributions are enhanced and thus the neutrino mode,
$p \rightarrow K^+ \bar{\nu}$, dominates.

\subsection{Decay Rates}
Next we will derive constraints on the charge assignment of the MSSM fields
from the null results of the  proton decay searches.
In the framework we are considering, the decay rates for the modes 
$p \rightarrow K^+ \bar{\nu}, K^0 e^+,K^0 \mu^+$ and 
$n \rightarrow K^0 \bar{\nu}$ can be comparable in  magnitudes.
However, since the most severe experimental constraint comes from the 
decay mode 
$p \rightarrow K^+ \bar{\nu}$, all we have to do is to study this mode.
 
The partial decay rate for the mode $p \rightarrow K^+ \bar{\nu}$ is 
calculated as
\begin{eqnarray}
  \Gamma(p \rightarrow K^+ \bar{\nu}) \sim \frac{m_p}{32 \pi} 
  \left( 1 - \frac{m_{K^+}^2}{m_p^2} \right)^2 \frac{1}{f_\pi^2} 
  \left( \sum_k |R_L \beta_p Amp_{Lk} + R_R \alpha_p Amp_R \delta_{k3}|^2 
  \right),
\end{eqnarray}
where
\begin{eqnarray}
  Amp_{Lk} & \sim & \frac{f}{M} 
  g_2^2 y_t y_b \lambda^{- h_u - h_d + \delta + 8 + (l_k - l_3)}, \\
  Amp_R & \sim & \frac{f}{M} 
  g_2^2 y_t y_b \lambda^{- h_u - h_d - \delta + 5+p} 
\frac{\tan \beta}{2}.
\end{eqnarray}
Here, $m_p$ and $m_{K^+}$ are the masses of proton and $K^+$ respectively,
$f_\pi$ is the pion decay constant, $R_{L,R}$ represent renormalization effects
of LLLL and RRRR operators from $M$ to $1~\mbox{GeV}$, and $\alpha_p$ and 
$\beta_p$ are the hadronic matrix element parameters.
Hereafter, for simplicity we fix the parameters as 
$m_{\rm SUSY} = 1~\mbox{TeV}, |\alpha_p| = 0.015~\mbox{GeV}^3, |\beta_p| = 0.014~\mbox{GeV}^3 \cite{Aoki:2000tw}, R_L = 10.2$ and $R_R = 6.5.$
\footnote{
$R_{L,R}$ are evaluated by solving one-loop renormalization group equations
due to gauge interactions
from a high energy scale ($\sim 2 \times 10^{16} ~\mbox{GeV} $) 
to $1~\mbox{GeV}$. 
Inclusion of the Yukawa couplings and 
effects above this high energy scale
do not qualitatively change our results.}

For the {\em best-fit} charge assignment, in order not to conflict 
with the experimental bound 
$\tau(p \rightarrow K^+ \bar{\nu}) > 1.9 \times 10^{33}~\mbox{yr}$
\cite{SK},
the U(1) charges must satisfy the following inequalities:
\begin{eqnarray}
  y_t y_b \lambda^{- h_u - h_d + \delta + 8} 
  & \lesssim & 4.2 \times 10^{-8}, \\
  y_t y_b \lambda^{- h_u - h_d - \delta + 10} \frac{\tan \beta}{2} 
  & \lesssim & 6.2 \times 10^{-8}.
\end{eqnarray}
Substituting the approximate values 
$m_t \sim 110~\mbox{GeV}$ and $m_b \sim 1.0~\mbox{GeV}$ 
near the Planck scale, the above  are rewritten as
\begin{eqnarray}
  \lambda^{- h_u - h_d + \delta} \tan \beta & \lesssim & 2.1,\\
  \lambda^{- h_u - h_d - \delta} \frac{\tan^2 \beta}{2}
  & \lesssim & 64.
\end{eqnarray}
Thus, we obtain
\begin{eqnarray}
  h_u + h_d & \lesssim & - 0.2 + \delta - 0.7 \ln (\tan \beta/3),\\
  h_u + h_d & \lesssim & 1.8 - \delta - 1.3 \ln (\tan \beta/3).
\end{eqnarray}
The proton decay constraint is  relaxed most when  
$\tan \beta \sim 3$ and 
$\delta=1$, which gives the upper bound 
\begin{eqnarray}
  h_u + h_d \lesssim 1,
\end{eqnarray}
and hence we find that the case $h_u + h_d=0$
survives the proton decay constraint for very low $\tan \beta$.
On the other hand, for larger $\tan \beta$, the sum of the two Higgs charges
must be negative for any choice of $\delta$ to satisfy the proton decay bound.

For the GUT-inspired case where $\delta = 0$ and $p= 3$,
the constraints on the Higgs charges are modified as
\begin{eqnarray}
  h_u + h_d & \lesssim & - 0.2 - 0.7 \ln (\tan \beta/3),\\
  h_u + h_d & \lesssim & - 0.3 - 1.3 \ln (\tan \beta/3).
\end{eqnarray}
Enhancement of the RRRR contributions arises from the fact that
difference between $u_1^c$ and $u_3^c$ is smaller than that in the 
{\it best-fit} case.
Thus we  conclude that
the charge assignment $h_u + h_d=0$ is marginally allowed for very small 
$\tan \beta$.

In deriving the constraints above, one has to keep in mind that there are 
some uncertainties in evaluating the proton decay rate. They include
\begin{enumerate}
\item The coefficients of the proton-decay dimension five operators which have
been set order unity may be accidentally small,  like $m_u$ in the 
GUT-inspired case. 
\item There are some uncertainties in 
  $\alpha_p$ and $\beta_p$. Taking the smallest allowed values for them,
  the proton decay amplitudes become five times smaller than our estimate.
\item We have taken the representative sparticle mass $m_{SUSY}=1$
  TeV. However, it becomes larger and the proton decay gets suppressed
  if we adopt smaller wino and higgsino masses and larger squark and
  slepton masses.
\end{enumerate}
These could reduce the proton decay rate, and then
the constraints for the Higgs charges obtained above would 
become somewhat relaxed. 

On the other hand, negative charge assignment 
for the Higgs doublets \cite{Maekawa:2001uk} 
survives the proton decay constraint in a wider region of the parameter space 
as the proton decay rate is reduced by $\lambda^{-2(h_u+h_d)}$.

\section{$\mu$ Parameter}
Here we would like to make a brief comment on the higgsino mixing
parameter $\mu$ when $h_u + h_d$ is negative. In this case, neither
the term $[\mu H_u H_d]_{\theta^2}$ nor $[\mu' (X/M_{\rm pl})^n H_u
H_d]_{\theta^2}$ (with $n>0$) is invariant under the Froggatt-Nielsen
U(1) symmetry. Thus
the generation of the $\mu$ term is somewhat contrived. Consider the
Giudice-Masiero mechanism \cite{Giudice:1988yz} with the term
$[Z^{\dagger}/M_{\rm pl}(X^{\dagger}/M_{\rm pl})^{-(h_u+h_d)}H_u
H_d]_{\theta^2 \bar \theta^2}$. This gives the $\mu$ term of
$F_Z^{\dagger}/M_{\rm pl} \lambda^{-(h_u+h_d)}\sim m_{3/2}
\lambda^{|h_u+h_d|}$.  This is much smaller than the electroweak scale
unless the gravitino mass $m_{3/2}$ is large. This is, in fact, the
case in, {\it e.g.} anomaly mediation where $m_{3/2}$ is around a few
tens TeV. Thus as far as $\lambda^{|h_u+h_d|}\gtrsim 10^{-2}$, we
obtain the weak scale $\mu$ parameter.  However, the Higgs mixing
parameter, $B$, is not suppressed and in general as heavy as
$m_{3/2}$. Then we will have difficulty in obtaining the correct
electroweak symmetry breaking.

A possible mechanism is to introduce a singlet $S$ with a positive U(1)
 charge so that $[S H_u H_d]_{\theta^2}$ is invariant under
the flavor symmetry. It is assumed that $S$ field does not have a VEV
in the SUSY limit. SUSY breaking generally generates a VEV of $S$,
which is nothing but the $\mu$ \cite{Kawamura:1995ys}. 
An explicit realization was given in
Ref. \cite{Maekawa:2001yh}, where the charges of relevant fields should be
adjusted to obtain the $\mu$ parameter of correct order of
magnitude. Thus the choice of a negative $h_u + h_d$ does not
immediately cause trouble to generate the $\mu$ parameter, though the
way to do is rather restricted.

\section{Conclusions and discussion}

In this paper, we have considered the proton decay when the dimension five
operators are controlled by the U(1) flavor symmetry.  We have shown 
that the charged lepton decay modes can have sizable branching ratios
when the LLLL operators dominate over the RRRR ones.  
In this case, measurement of
the branching ratio of the $e$ mode and the $\mu$ mode will provide
an important information on  the charges for the doublet leptons and thus
reveal how the neutrino masses are generated. On the other hand, if the RRRR 
operators are the main sources of the proton decay, the neutrino mode dominates
in the proton decay.  We have also investigated
the proton decay rate and showed that the proton-decay suppression due
to the U(1) flavor symmetry marginally survives the present proton decay
bound even when the Higgs doublets do not have the U(1) charges. The
bound is less severe, if we take negative charges of the Higgs doublets.

We have made several assumptions in deriving the conclusions drawn above.
Firstly we have assumed that non-renormalizable operators are suppressed
by the four-dimensional Planck scale. However, there are many models in which
the fundamental scale is lower than the Planck scale. In this case, it is
natural to expect that the proton-decay dimension five operators are suppressed
only by this low fundamental scale and hence  the proton
decay rate is considerably enhanced.
 To survive the present experimental bound, one needs to
assign the negative U(1) charges to the Higgs doublets in this case. 
Arguments on the branching ratios remain unchanged.

Another important assumption we have made is the degeneracy of the squark 
and slepton masses in different generations. 
Let us consider the case where
the superparticles of the first two generations are heavy, which is
realized when an anomalous U(1) flavor symmetry also mediates SUSY breaking. 
In this scheme, as far as the scalar tops $\tilde{t}_{L,R}$ and the 
right-handed scalar tau $\tilde{\tau}_R$ are light like in the case of GUTs,
the amplitudes of the LLLL operators are greatly suppressed
while those of the RRRR operators remain unsuppressed
\cite{Kakizaki,Murayama:2002ur}.
Thus, one finds that $p \rightarrow K^+ \bar{\nu}$ tends to dominate over 
$K^0 e^+,K^0 \mu^+$ modes.
On the contrary, when only $\tilde{t}_{L,R}$ are light, 
the RRRR operators are suppressed to the same extent of the LLLL operators.
Therefore, the discussions about the branching ratios are not altered.

\section*{Acknowledgment}               
This work was supported in part by the
Grant-in-aid from the Ministry of Education, Culture, Sports, Science
and Technology, Japan, priority area (\#707) ``Supersymmetry and
unified theory of elementary particles,'' and in part by the
Grants-in-aid No.11640246 and No.12047201.

\end{document}